# Self-Supervised Deep Learning to Enhance Breast Cancer Detection on Screening Mammography

John D. Miller, Vignesh A. Arasu, Albert X. Pu, Laurie R. Margolies, Weiva Sieh, Li Shen

*Abstract* — A major limitation in applying deep learning to artificial intelligence (AI) systems is the scarcity of high-quality curated datasets. We investigate strong augmentation based self-supervised learning (SSL) techniques to address this problem. Using breast cancer detection as an example, we first identify a mammogram-specific transformation paradigm and then systematically compare four recent SSL methods representing a diversity of approaches. We develop a method to convert a pretrained model from making predictions on uniformly tiled patches to whole images, and an attention-based pooling method that improves the classification performance. We found that the best SSL model substantially outperformed the baseline supervised model. The best SSL model also improved the data efficiency of sample labeling by nearly 4-fold and was highly transferrable from one dataset to another. SSL represents a major breakthrough in computer vision and may help the AI for medical imaging field to shift away from supervised learning and dependency on scarce labels.

## I. Introduction

AS the field of computer vision matures and produces increasingly powerful models, interest in their application to medical imaging data has grown [1]. One important application is the detection of breast cancer on screening mammograms. Breast cancer is the most common cancer and the second leading cause of cancer deaths among US women [2]. Screening mammography has been shown to significantly improve outcomes and reduce mortality [3]. Early computer-aided detection (CAD) software performed poorly and did not increase the accuracy of radiologists [4, 5]. More recently, several large-scale studies [6-9] found that newer AI for mammography systems, boosted by the adoption of deep learning, have performed on par with human readers, and often improved performance when combined with radiologists' decisions.

While such developments are encouraging, creating a large curated medical imaging dataset to train deep learning algorithms is both expensive and time-consuming. Unlike photographic images, which can be obtained on the Internet and labeled by non-experts through crowd-sourcing, radiographic image curation requires domain knowledge. Hence, even prior studies with abundant resources used training sets that are dwarfed by the ImageNet [10] in sample size. Additionally, deep learning models are data hungry and continue to improve as they are fed with more data [11].

Many techniques have been developed to address the issue of label scarcity in medical imaging [12], which can be categorized into three major groups: semi-supervised learning, multiple instance learning, and transfer learning. Recently, self-supervised learning (SSL) emerged as a combination of semi-supervised and transfer learning methods. Early SSL methods used pretext tasks to pretrain a model and showed promise in a range of vision tasks [13]. More recent iterations of SSL methods, using strong augmentations such as cropping, color distortion and masking, started to show results that approached fully supervised state-of-the-art models [14-19]. Fundamentally, such SSL methods seek to pretrain an encoder network using strongly augmented views of unlabeled images to extract features for a self-supervised task. The encoder network can then be used in a variety of downstream tasks, such as image classification and object detection. Because the pretraining can potentially utilize a large quantity of unlabeled data to produce a strong encoder, the model may achieve high performance even when the supervision is limited.

Because strong augmentation based SSL represents the latest development in computer vision, they have received little attention in medical imaging studies to date [20-22]. In this study, we attempt to develop an SSL based method for breast cancer detection using screening mammograms. Our contributions are multifold. First, we systematically evaluate the transformations that are tailored to the X-ray images, which allow the SSL pretraining to be effective. Second, we illustrate a patch to whole-image training scheme that achieves excellent classification performance while meeting the computational requirements of SSL. Third, we show that whole-image classification can benefit from attention-based pooling. Fourth, we benchmark several recent SSL methods for both patch and whole-image classification. Finally, we demonstrate that the pretrained models are transferrable across two datasets with different intensity profiles.

Icahn School of Medicine at Mount Sinai, New York, NY (J.D.M., L.R.M., W.S., L.S.); Kaiser Permanente Northern California, Division of Research, Oakland, CA (V.A.A., A.X.P.). Address correspondence to: li.shen@mssm.edu.

## II. METHODS

### A. Self-Supervised Learning (SSL) Based on Strong Augmentations

We selected four representative SSL methods that are unique in their formulation of the self-supervised tasks. These models are based on neural networks optimized using stochastic gradient descent (SGD) methods on training samples that are randomly assigned to "batches". Here, we provide a brief introduction to each method.

**SimCLR** [18] is the first SSL method that uses strong augmentation to attain performance levels near supervised learning methods. It encodes two augmented images ("views") of the original image using the same encoder network and tries to maximize the agreement between the two augmented views while minimizing the agreement between each augmented view and the other views in the same batch.

**Bootstrap your own Latent (BYOL)** [15] tries to maximize the agreement between one view that is encoded by an online network and a different view encoded by a target network. It does not use the other views in the same batch as a contrast.

**Swapping Assignments between Views (SWaV)** [14] assigns one view to one of the learned prototypes to generate a code, while calculating the soft assignment for the opposite view. It tries to maximize the agreement between the code and the soft assignment through a cross-entropy loss. The prototypes are like cluster centroids that are continuously updated throughout training.

**Visual Transformer Masked Autoencoder (ViT-MAE)** [16] performs a different kind of augmentation from the above methods by splitting the original image into patches, heavily masking a significant portion of them and using the remaining patches to reconstruct the input through a visual transformer (ViT) [23].

### B. Whole Image Classification with Tiled Patches and Attention Pooling

A classic way to test an SSL method is to train an encoder and then evaluate a linear classifier using the features extracted by the encoder with its parameters fixed, known as linear evaluation. Initially, we tested the SSL methods directly on the whole mammographic images but the performance did not surpass an ImageNet pretrained model. The reasons behind this may be two-fold. First, some augmentation methods, such as cropping, are not well suited for medical image classification because the key areas, such as cancer lesions, are localized. Second, a large training batch size is an important factor in the success of many SSL methods [15, 18]. Mammograms are very large, restricting the batch size that can feasibly be analyzed on available computational resources (we had up to 8 GPUs on our server). To address this limitation, we split the whole images into patches, pretrained the models on the patches, and then converted them into whole-image models in the same way as in our previous study [24]. A diagram of this paradigm is shown in Figure 1.

After an encoder network is pretrained on patches, it is applied on a whole image to produce a grid of embeddings $H \in \mathbb{R}^{r \times c \times m}$ where $r$ and $c$ are the spatial dimensions of the grid and $m$ is the embedding size. We can rearrange $H$ so that $H = \{h_1, \ldots, h_K\}$ is a bag of embeddings, where each $h_k \in \mathbb{R}^m$ and $K = rc$. An appropriate pooling method is needed to condense $H$ into a representative feature vector for the entire image. An obvious choice is global-average-pooling (GAP) where the embeddings are simply averaged:

$$z = \frac{1}{K} \sum_{k=1}^{K} h_k \quad (1)$$

where $z$ is used for linear evaluation.

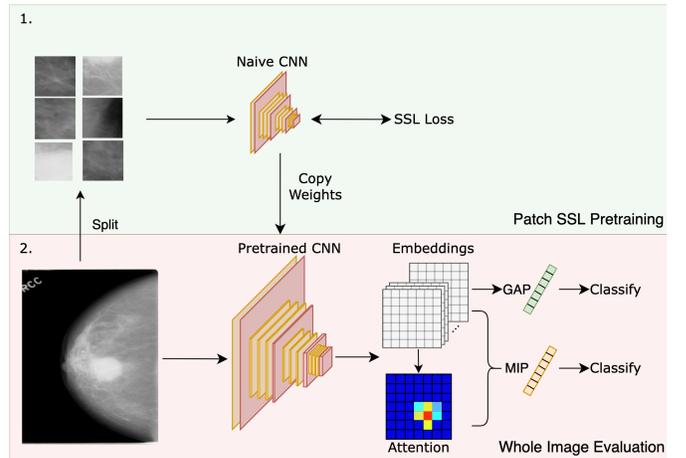

Figure 1. Tiled-patch pretraining to whole image method. Models are first pretrained using tiled patches before being adapted to evaluate whole mammograms.

A potential problem with GAP is that it may dilute the signal because only few of the embeddings are relevant to classification. Inspired by Illse et al. [25], we also investigated Multiple-Instance-Pooling (MIP) that is based on an attention mechanism. MIP utilizes a light-weight two-layer network to calculate an attention score for each $h_k$, allowing them to be pooled as a weighted average. The attention for $h_k$ is calculated as:

$$a_k = \frac{\exp\{w^T(\tanh(Vh_k) \odot \mathrm{sigm}(Uh_k))\}}{\sum_{i=1}^{K} \exp\{w^T(\tanh(Vh_i) \odot \mathrm{sigm}(Uh_i))\}} \quad (2)$$

where $V \in \mathbb{R}^{n \times m}, U \in \mathbb{R}^{n \times m}, w \in \mathbb{R}^n$ are learnable weights; tanh and sigm represent hyperbolic tangent and sigmoid activations, respectively; and $\odot$ represents the element-wise product. For the linear layers, we used $n = 512$ across the board. For $m$, it depends on the network structure. For a standard ResNet-50, $m = 2048$. Note that the attention scores are fed through a softmax activation so that they sum up to one. Using the attention scores, the embeddings can then be pooled by a weighted average:

$$z = \sum_{k=1}^{K} a_k h_k \quad (3)$$

It is important to note that MIP is permutation invariant. In other words, the $a_k$ are determined independently from other embeddings except for the softmax squashing effect. We found such an attention mechanism to be sufficient for mammograms. Additionally, ViT-like self-attention (SA) can be used, which is explored in Alternative Designs.

## III. DATASETS

Two datasets were used to test the SSL methods: the Curated Breast Imaging Subset (CBIS-DDSM [26]) of the DDSM [27] collection of digitized screen-film mammograms across four sites in the United States; and the Chinese Mammography Database (CMMD) [28] of full-field digital mammograms (FFDMs) collected by researchers at South China University of Technology. Both datasets were attained through The Cancer Imaging Archive (TCIA) Public Access [29]. Although both datasets contain multiple images from one exam for each patient, we treated each image as a separate sample and constructed training and evaluation sets by splitting the data at the patient level so that all images from the same patient are in the same set. The sample distributions for training-validation-test images for both datasets are shown in Supplemental Fig. 1.

### A. CBIS-DDSM

The CBIS-DDSM contains 3119 scanned film mammograms from 1566 participants and includes both craniocaudal (CC) and mediolateral oblique (MLO) views. All 761 cancer cases were pathologically confirmed with regions of interest (ROIs) annotated. At the patient level, the dataset was split approximately 7:1:2 into training, validation and test sets so that they contained 2111, 367, and 641 images, respectively. The splits were stratified to maintain the same proportion of cancer cases in all subsets. The images were downsized to 1152x896 and converted into 16-bit PNG files.

We used the ROIs to create an annotated patch set that contained one patch centered on an ROI, and one drawn at random from the rest of the breast region per mammogram. Patches were drawn at 224x224 pixels and stored as 16-bit PNG files. All patches were classified as one of five categories: background, malignant mass, benign mass, malignant calcification, and benign calcification based on the annotation of ROIs. Sample distributions of the patches can be found in Supplemental Fig. 1.

### B. CMMD

The CMMD is composed of 3728 FFDMs from 1775 patients and includes both MLO and CC views. All patients in the dataset have malignant or benign tumors, with malignancies confirmed by biopsy. No ROI-level annotation is included. The mammograms were split into training, validation, and test sets at a 3:1:1 ratio at the patient level. The splits were stratified to maintain the same proportion of malignant cases. This resulted in training, validation and test sets containing 1065, 355 and 355 patients with 2237, 745, and 746 images, respectively. Images were downsized from their original size of 2294x1914 to 1147x957 and saved as 16-bit PNG files.

### C. Tiled Patch Set Creation

As described above, we used the whole images to generate tiled patch sets from CBIS-DDSM and CMMD for pretraining. Patches were drawn in a grid pattern over the whole image with 50% overlap. We allowed overlap amongst the patches to avoid arbitrarily drawn boundaries on lesions. Any patches that contained more than 20% background were removed because too much breast boundary information may allow a network to take advantage of the shape and bypass the effort to recognize texture patterns in the patches. Note that the tiled patches did not contain any labels. We drew patches at sizes 64x64, 96x96, 128x128, and 256x256 pixels to find out which patch size worked the best.

## IV. ANNOTATED PATCH SET CLASSIFICATION

As applying SSL directly on the whole images failed to yield meaningful results initially, we sought to test the SSL models on the annotated patch set as a proof of concept. We also used the annotated patch set to evaluate which transformations would work best for mammographic images. These efforts involved pretraining directly on the annotated patches and classifying them into one of the five categories using linear evaluation. Note that such patch-level annotation is uncommon as it requires significant curation, and is not expected to be available for most large mammography datasets.

### A. Mammogram-Specific Augmentations

Strong data augmentation has been shown to be essential for effective SSL [18]. Many SSL methods have relied on a shared transformation paradigm that is typically composed of cropping, color distortion and Gaussian blur [14, 15, 18, 19]. However, this transformation scheme has been optimized to maximize performance for photographic color images. Mammograms are grayscale and contain very different features. To address this, we need to develop a mammogram-specific augmentation method.

We examined eight transformations that altered the images both spatially and in appearance: random crop with a resize to the original image size, brightness shift, contrast shift, gamma shift, Gaussian blur, histogram equalization, sharpening, and high-pass filter. Examples of the eight transformations are shown in Fig. 2. Additionally, we duplicated the patches onto three channels to make the input compatible with most standard convolutional neural networks (CNNs) for vision. Note that the transformations were applied to the images before casting them onto the three channels so as not to add pseudo-colors.

The transformations were evaluated both alone and in a pairwise fashion on the annotated patch set. We chose BYOL to conduct this experiment because it is easy to implement, relatively insensitive to batch size, and performed well in our initial tests. Specifically, BYOL was pretrained with a single or pair of transformations applied to both views for 200 epochs at a batch size of 2048 across 4 GPU cores. The LARS optimizer [30] was used at a base learning rate of 0.4 and global weight decay of $1.5 \times 10^{-6}$. The LARS applies a layer-wise learning rate adaptation throughout training to help with large batch size. Both biases and batch normalization parameters were excluded from such adaptation and weight decay. The pretrained models were then linearly evaluated at a learning rate of $10^{-3}$ and weight decay of 0.01 with the Adam optimizer [31].

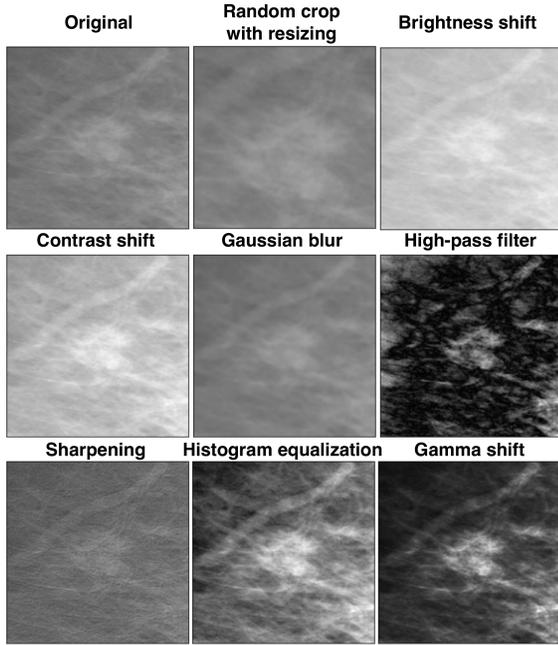

Figure 2. Sample images for the 8 transformations tested: random crop with resizing, brightness shift, contrast shift, Gaussian blur, high-pass filter, sharpening, histogram equalization, and gamma shift.

Fig. 3A shows the results of linear evaluation on models trained with all 64 combinations of single or pairwise transformations. Most notably, cropping proved to be essential to model performance. No pair of transformations performed strongly in the absence of cropping. Additionally, the matrix is mostly symmetric, suggesting that the order of transformations is unimportant to pretraining. Due to this, the four best performing transformations were selected and tested again pairwise in tandem with cropping (Fig. 3B). Of these, a shift in both gamma and contrast and histogram equalization proved to be the strongest. We chose crop, gamma, and contrast transformations for further experiments in this study because histogram equalization is more computationally demanding.

### B. SSL Model Comparison on Annotated Patch Classification

Once the optimal mammogram augmentation paradigm was established, we compared the four SSL methods using pretraining and linear evaluation on the annotated patch set. As ViT-MAE uses masking, no transformation was used for this method. Models were pretrained for 200 epochs with hyperparameters following each model's original publication before being linearly evaluated for an additional 100 epochs at a learning rate of $10^{-3}$ and weight decay of 0.01. SWaV's performance was somewhat sensitive to the number of prototypes and the queue length. We tested several combinations and decided to use 50 prototypes with a queue length of 300. Table 1 shows that BYOL outperformed all the other models, SWaV was a close second, while SimCLR and ViT-MAE performed much worse. Surprisingly, ViT-MAE's performance was extremely poor compared with the other models, suggesting that ViT based models may be poorly suited for mammographic image classification tasks. Unlike natural images, mammograms do not contain the shapes and/or texture patterns for parts of the image to infer the state of the whole image. This may have contributed to the MAE's inability to learn features from masked mammographic images.

Figure 3. Linear evaluation accuracy % on the annotated patch test set. (A) Pairwise transformations for all 8 transformations. Cells along the diagonal are single transformations. (B) A second pairwise test, this time always including a crop as the first transformation.

Table 1. Test set accuracies of different SSL methods on the annotated patch dataset.

| Model | Test Set Accuracy |
| --- | --- |
| *SimCLR* | 0.653 |
| *SWaV* | 0.741 |
| **BYOL** | **0.763** |
| *ViT-MAE* | 0.550 |

## V. CBIS-DDSM TILED PATCH PRETRAINING TO WHOLE IMAGE CLASSIFICATION

Next, we sought to transfer our findings from annotated patches to whole mammograms without relying on the detailed annotation from CBIS-DDSM. We pretrained the models on tiled patches uniformly drawn from whole mammograms and then evaluated them on the whole images. Note that the tiled patches did not contain any labels. Only the whole images were labeled as positive, indicating the presence of a malignant lesion, or negative, indicating the presence of a benign lesion or absence of any abnormalities.

## A. Patch Size Selection

To investigate which patch size can produce the strongest results on whole image classification, BYOL was used for pretraining on separate tiled patch sets of differing sizes with a batch size of 4096. The LARS [30] optimizer was used at a learning rate of 4.8 with weight decay of $1.5 \times 10^{-6}$. Due to memory constraints, the batch size was reduced to 2048 and learning rate accordingly adjusted to 2.4 in analyses of 256x256 size patches. Biases and batch normalization parameters were excluded from weight decay and LARS adaptation. The best model was chosen based on the pretraining loss on the validation set, which correlates well with whole image classification performance (Supplemental Fig. 2). The pretrained models were then linearly evaluated on the whole images for 100 epochs at a learning rate of $10^{-4}$ and weight decay of 0.001 with the Adam optimizer [45] using GAP. Table 2 shows classification AUCs on the CBIS-DDSM whole image test set. Interestingly, the best performing patch size was 96x96, although 256x256 performed similarly. However, given the batch size limitation using 256x256 patches and the batch size being a contributing factor to SSL model performance, 96x96 patches were used in further experiments.

## B. SSL Model Comparison on Whole Image Classification

With the tiled patch size determined, we returned to the SSL model comparison on whole images. BYOL, SimCLR and SWaV are all based on CNNs, making them natural to convert from patch to whole image. ViT-MAE works by encoding the unmasked patches for the ViT, so splitting the whole image into patches beforehand was unnecessary. Additionally, ViT expects the pretraining and evaluation input size to be the same, making it difficult for the patch-to-whole-image conversion.

The three CNN-based models were pretrained for 200 epochs at a batch size of 4096 split across four NVIDIA A100 GPUs. SimCLR was pretrained using the LARS optimizer at a learning rate of 4.8 and weight decay of $1.5 \times 10^{-6}$ with biases and batch normalization excluded from LARS adaptation and weight decay. The temperature was set to 0.5. Batch normalization layers were synchronized across GPUs. BYOL and SWaV were pretrained under the same paradigm. BYOL's EMA decay rate was set to be 0.99. SWaV's temperature parameter was set to 0.1 and the Sinkhorn regularization parameter was set to 0.05. We used 20 prototypes and a queue length of 3000 when training SWaV. ViT-MAE was pretrained for 200 epochs at a batch size of 32 on the same hardware using the AdamW [32] optimizer at a learning rate of 0.0024 and weight decay of 0.05 on the whole images.

The pretrained models were frozen and linearly evaluated on the whole images using GAP for 100 epochs at a batch size of 32. Training was performed using the Adam optimizer at a learning rate of $10^{-4}$ and weight decay of 0.001. These hyperparameters were the same regardless of the pretraining method. Interestingly, SWaV performed the strongest in linear evaluation on whole mammograms (Table 3), although it fell behind BYOL on the annotated patch set (Table 2). It is important to note that the tiled patch set is more imbalanced than the annotated patch set with a lot of background patches. This may favor clustering-based approaches, such as SWaV, over instance-based approaches, such as SimCLR and BYOL. ViT-MAE continued to underperform compared with the other methods.

Table 2. Test set AUCs on whole mammogram classification using BYOL with GAP on different patch sizes.

| Patch Size | Test Set AUC |
|---|---|
| 64x64 | 0.598 |
| **96x96** | **0.687** |
| 128x128 | 0.642 |
| 256x256 | 0.682 |

Table 3. Test set AUCs of whole mammogram linear evaluation and finetuning with different pretraining and pooling methods.

| Model | Linear Evaluation | Finetuning |
|---|---|---|
| *SimCLR (GAP)* | 0.674 | - |
| *ViT-MAE* | 0.600 | - |
| *BYOL (GAP)* | 0.687 | 0.763 |
| *BYOL (MIP)* | 0.722 | 0.783 |
| *BYOL (SA)* | 0.691 | 0.744 |
| **SWaV (GAP)** | **0.709** | **0.803** |
| **SWaV (MIP)** | **0.757** | **0.815** |
| *Supervised* | 0.710 | |

Thus far, we have conducted all whole image experiments using GAP. Due to a lesion generally being contained within a local area of a mammogram, GAP may wash out valuable information contained in this region's embedding. To counter this, we replaced the GAP layer with a MIP layer for the SWaV and BYOL pretrained models. The evaluation scheme was kept the same from previous whole image experiments. MIP exceeded the performance of GAP as much as 0.05 for both models (Table 3). We further show the attention heatmaps for SWaV using a few examples (Fig. 4). Intense focus can be seen on potential lesions in the sample heatmaps. This suggests that the SWaV pretrained model can provide patch-level features for the MIP layer to learn to focus on the lesions that are most important for the whole image classification. Such focus leads to the improved quality of the pooled feature vector and therefore the increased classification performance. Similar heatmaps for BYOL can be found in Supplemental Fig. 3.

## C. Model Finetuning

One way to utilize SSL for mammogram classification is to pretrain a model using a large number of unlabeled images and later finetune the model with a much smaller labeled dataset. To simulate this scenario, we finetuned the BYOL and SWaV pretrained models from above on the CBIS-DDSM whole images using the following strategy:

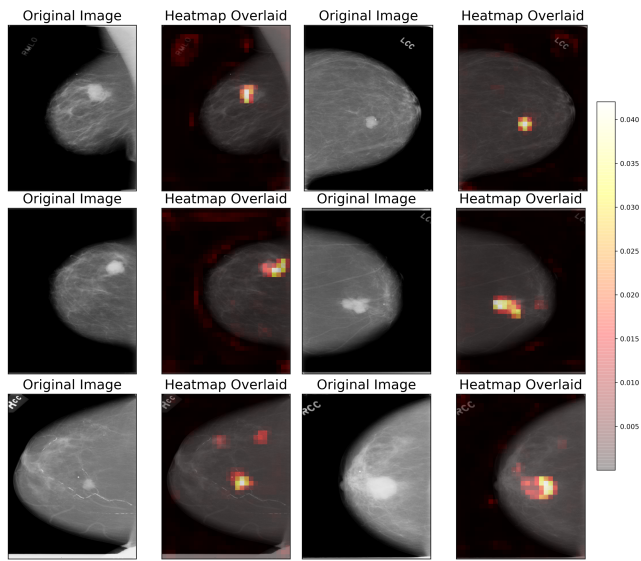

Figure 4. Sample heatmaps from 6 mammograms in the training set. An MIP attention layer was trained on top of a frozen ResNet-50 encoder pretrained using SWaV on 96x96 tiled patches. The color bar corresponds to probabilities generated by the softmax layer of MIP. Larger numbers indicate a larger assigned weight.

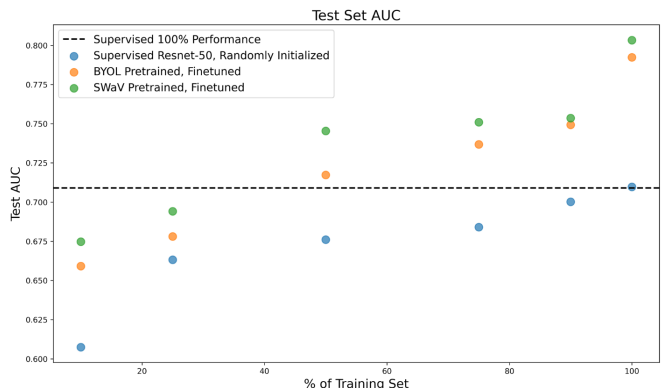

Figure 5. Supervised vs SSL-pretrained performance by subset of the training dataset.

1. Freeze the pretrained model, and train only the final linear layer for 100 epochs at a learning rate of $10^{-4}$ and weight decay of 0.001. Images are not augmented for this step. This is the same as our previous linear evaluation protocol.
2. Unfreeze the remainder of the model and train for an additional 50 epochs at a learning rate of $10^{-5}$ and weight decay of 0.01. Images are augmented with a random resized crop between 80 and 120% of the original image's size, a random horizontal and vertical flip and rotation in [-25, 25] degrees, and a random brightness adjustment in [-20, 20] in pixel values.

For comparison, a ResNet-50 model (baseline) was trained from scratch in a supervised manner for 200 epochs at a learning rate of $10^{-4}$ with no weight decay and the same image augmentations used as in step 2 above. Note that a different set of data augmentations was used in finetuning because we observed a performance drop when the same data augmentations as SSL were used. All models were trained using a 10, 25, 50, 75 and 90% subset of the training data, along with the entire set (Fig. 5).

Table 4. Evaluation and finetuning of ResNet-152 and ResNet-50-2x. Shown are CBIS-DDSM whole image test set AUCs. Both models were evaluated with both GAP and MIP.

| Model | Linear Evaluation | Finetuning |
|---|---|---|
| *ResNet-50-2x (GAP)* | 0.703 | 0.803 |
| *ResNet-152 (GAP)* | 0.703 | 0.797 |
| *ResNet-50-2x (MIP)* | 0.751 | 0.782 |
| *ResNet-152 (MIP)* | 0.755 | 0.807 |

The finetuned models outperform the supervised baseline in every subset, demonstrating significantly improved data efficiency. In fact, even with 100% of the training set, the supervised baseline barely exceeded the performance seen by the SWaV-pretrained model with only 25% of the training data (Fig. 5). This reinforces the main strength of self-supervised pretraining: it can greatly reduce the costs associated with generating datasets going forward, as a much smaller annotated training set is required.

Additionally, we performed finetuning for both SWaV and BYOL pretrained models with MIP. MIP also benefits from finetuning, albeit to a lesser degree than GAP (Table 3). This is likely due to the high performance of MIP diminishing the gains from finetuning.

### D. Alternative Designs

Studies on SSL methods have often demonstrated a significant boost in performance when replacing the image encoder with a bigger model, such as a wider or deeper ResNet [15, 17, 18]. This is intuitive – SSL is unsupervised in nature and requires a larger model to incorporate a more comprehensive set of features. To investigate this possibility, we pretrained both a deeper and a wider ResNet using the SWaV. For a deeper model, we selected ResNet-152, which contains 3x as many convolutional layers as ResNet-50 [33]. For a wider model, we chose ResNet-50-2x, which has the same structure as the ResNet-50 but twice the number of filters per convolutional layer [34]. Pretraining was done with the same scheme described previously. These models were then evaluated and fine-tuned with both GAP and MIP. Surprisingly, there is no improvement in the use of larger models (Table 4). This may be due to the relative simplicity of mammographic images in comparison to natural images that contain a more diverse set of features.

SA layers such as the ones used in ViT [23] have the potential to improve classification by incorporating inter-relationship among patches. To investigate this, we replaced the GAP layer of a ResNet-50 with an SA layer in BYOL pretraining. In each SA layer, we prepended a classification token to our image embeddings and used this to perform linear evaluation and finetuning. The addition of SA did not improve the model's classification performance (Table 3). Additionally, SA underperformed MAP in both linear evaluation and finetuning. This underwhelming performance of SA further supports the notion that SA based models are poorly suited to mammogram classification.

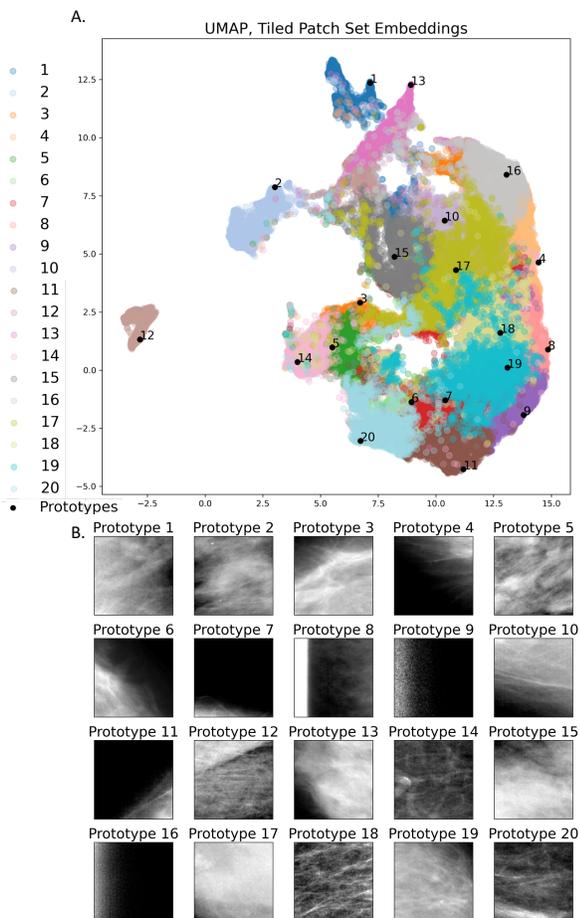

Figure 6. (A) UMAP projections of 50,000 96x96 patch embeddings. The prototypes are shown as black dots and labeled by their assignment numbers. (B) Sample patches from each prototype.

### E. SWaV Prototype Analysis

Given that SWaV yields soft assignments for images in pretraining, we sought to understand the distribution of the patches within these clusters. Using our pretrained SWaV network, we embedded and assigned 50,000 random samples from the 96x96 tiled patch training set to the model's prototypes. This was done in an online fashion to mimic the samples' assignments during pretraining. The soft assignments generated by SWaV were collapsed into discrete assignments corresponding to the largest probability. We then reduced the dimensionality of the embeddings using the uniform manifold approximation and projection (UMAP) [35] method, allowing us to visualize the clustering of our embeddings in two-dimensional (2D) space. We also reduced the dimensionality of the prototypes themselves using the same projection as the embeddings.

Fig. 6A shows the 2D projections of the embeddings and their discrete assignments. Interestingly, patches with the same assignment are generally grouped together in the 2D space, showing that SWaV can effectively learn to embed the patches and cluster them based on their similarity. Fig. 6B shows the closest patches to each prototype, measured by the Euclidean distance in the embedded space. Some interesting patterns emerge from the figure. For example, some prototypes seem to represent patches extracted from the boundaries of the breast (6, 7, 9, 11, 16). Others represent regions of high density with potential masses (3, 5, 15, 17). Prototype 2 is a distinct cluster of patches that may contain masses and calcifications, while prototype 12 is another distinct cluster of patches that contain pectoral muscles. The prototypes with similar appearance also tend to be near each other in the UMAP projection.

To further understand the prototypes generated by SWaV, we assigned the annotated patches to the prototypes using the same process. Note that each annotated patch belongs to one of the five categories: background, malignant mass, benign mass, malignant calcification, and benign calcification. We then investigated the distributions of the five classes amongst the prototypes (Supplemental Fig. 4). Several prototypes, such as 3, 10, 12, 16, and 17, were assigned nearly solely background patches. These prototypes alone account for over half of the background patch assignments. Prototypes 13, 18, and 19 were assigned a large portion of malignant calcification patches. While prototypes 5, 6, 14, and 18 were enriched with malignant mass patches. Due to SWaV's equipartition constraint (i.e., each prototype has to be assigned the same amount of patches during pretraining), we do not expect any prototype to be solely enriched with a single class of non-background patches. However, the fact that each prototype has distinct enrichment patterns of the five classes shows that SWaV pretraining learned to distinguish lesions from background, calcifications from masses, and malignant from benign lesions, all without being provided with any supervision. Such capability to distinguish different types of patches can be directly utilized in downstream classification tasks.

We also looked at the distributions of the 20 prototypes among the five classes (Supplemental Fig. 5). As expected, the background class has a very different histogram from the four non-background classes. The two calcification classes also have different histograms from the two mass classes. Interestingly, the benign and malignant classes have very similar histograms for either the calcification or mass type, reflecting similarities in their appearance.

## VI. CMMD TRAINING AND TRANSFER LEARNING

To test the universal applicability of the SSL methods for mammograms, we applied the SWaV models on the CMMD, which has a different intensity profile from the CBIS-DDSM (Fig. 7). We used a 5-fold cross-validation (CV) strategy for training and evaluation. For each CV iteration, the dataset is split 3:1:1 into training:validation:test sets. Patches were drawn and trained in the same manner as on the CBIS-DDSM data. We also transferred the CBIS-DDSM pretrained models onto the CMMD with the two pooling methods through both linear evaluation and finetuning.

As shown in Table 5, there is no significant difference between the CBIS-DDSM and CMMD-pretrained models. Surprisingly, even with a frozen encoder (i.e., linear evaluation), the results are nearly identical, suggesting that the SSL models are highly transferrable between datasets. This is in stark contrast with models learned via supervision, which often require labeled samples for finetuning when the two datasets have different intensity profiles [24, 36, 37]. Furthermore, both models with either form of pooling surpassed the supervised baseline after finetuning, highlighting the benefits of pretraining. Indeed, even linear evaluation with either form of pooling performed on par or slightly better than the supervised baseline.

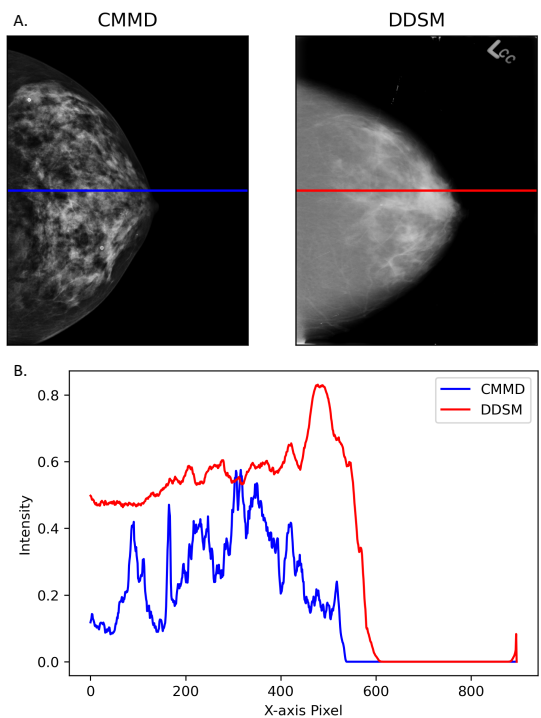

Figure 7. Comparison of intensity profiles between DDSM and CMMD. (A) Sample images and the lines drawn to gather intensity measures. (B) Intensity profiles of the two images.

Table 5: Mean evaluation and finetuning AUCs from 5-fold CV on the CMMD whole images. Standard errors are included in parentheses.

| Pretraining Set (Pooling Method) | Linear Evaluation | Finetuning |
|---|---|---|
| DDSM (GAP) | 0.702 (±0.005) | 0.734 (±0.006) |
| DDSM (MIP) | 0.714 (±0.007) | 0.730 (±0.007) |
| CMMD (GAP) | 0.703 (±0.006) | 0.732 (±0.006) |
| CMMD (MIP) | 0.712 (±0.007) | 0.730 (±0.007) |
| Supervised (GAP) | 0.706 (±0.013) | |

## VII. CONCLUSION & DISCUSSION

Here, we have developed an algorithm to adapt recent strong augmentation-based SSL methods for breast cancer detection on screening mammograms. SSL allows a model to learn to extract features on unlabeled samples and hence improves its performance in comparison to training from scratch. Specific transformations were identified to suit the mammography images. Pretraining on uniformly drawn patches instead of whole images made the training more computationally feasible yet still provided effective feature learning. Using a simple attention layer as a pooling layer significantly improved the classification performance in linear evaluation and to a less degree, in finetuning. Four diverse SSL approaches were systematically compared and SWaV, a clustering-based method, had the best performance. Finally, the model pretrained on one dataset was successfully transferred onto an independent dataset with different intensity profiles.

Interestingly, BYOL performed the best on annotated patch classification, while SWaV performed best on tiled-patch-to-whole-image classification. We postulate that this is due to the different distributions of patches in the two tasks. The tiled patches contain an overwhelming number of background images with a small proportion of abnormalities. Therefore, BYOL may be finding a shortcut to maximize the similarity between the two views of a patch. In contrast, clustering-based methods have innate capability to distinguish abnormalities from background. By enforcing an equal distribution of embeddings among a small number of prototypes, SWaV may be driving feature extraction towards a shared embedding of similar patches. Throughout pretraining, the model is optimizing its outputs to fit well on the prototypes, enforcing a stronger distinction between normal tissues and lesions.

Congruent with the results reported on ImageNet and other datasets, SimCLR performed worse than both BYOL and SWaV [14, 15]. This may indicate that SimCLR is a poorer method for feature learning on mammographic images. SimCLR's use of contrastive learning may also decrease performance when applied to tiled patches. Given that our dataset is composed of relatively homogenous patches, it may prove especially difficult to distinguish positive and negative pairs. This would lead to significantly weakened feature learning by the encoder network, ultimately harming its performance.

Also intriguing is the poor performance of the SA-based ViT-MAE. This suggests that SA mechanisms are poorly suited for medical images that contain small lesions amongst large quantities of background tissues, known as the "needle-in-a-haystack" problem. This may be due to the relative independence of a lesion from other surrounding tissues on the same image. SA relies heavily on the spatial relationships between patches of an image. However, if features of the background tissues have limited ability to predict the presence of a lesion, encoding the patch-to-patch correlations may be uninformative for whole image classification. Therefore, permutation invariant methods such as CNNs may be better suited for mammography imaging detection and diagnosis.

In summary, we have shown that our SSL method allows for the inclusion of significantly more unannotated data during pretraining to significantly boost the performance of mammography image classification. These study findings are important for informing future research because expansion of the training datasets readily available at major hospitals is critical for developing medical imaging algorithms that ultimately can improve patient care. The SWaV algorithm is especially promising based on its outstanding performance for both whole image classification and distinction of patches during unsupervised clustering. Future research is needed to investigate whether global clustering methods such as DeepCluster [38], or a hybrid of instance and clustering-based methods such as CLD [39], can further improve the identification and recognition of similar patches and distinction of dissimilar patches. This study provides a proof of concept that improvements in patch classification using unannotated data can substantially improve the performance of deep learning algorithms for the detection of breast cancer on screening mammography. In future work, we will examine extending our SSL method to digital breast tomosynthesis, which is rapidly increasing in use in the U.S.

ACKNOWLEDGEMENTS

This study was supported by grants from the National Institutes of Health (R01CA264987, R01CA237541) and by the computational resources and staff expertise provided by Scientific Computing at the Icahn School of Medicine at Mount Sinai.

# Supplemental Figures for "Self-Supervised Deep Learning to Enhance Breast Cancer Detection on Screening Mammography"

John D. Miller, Vignesh A. Arasu, Albert X. Pu, Laurie R. Margolies, Weiva Sieh, Li Shen

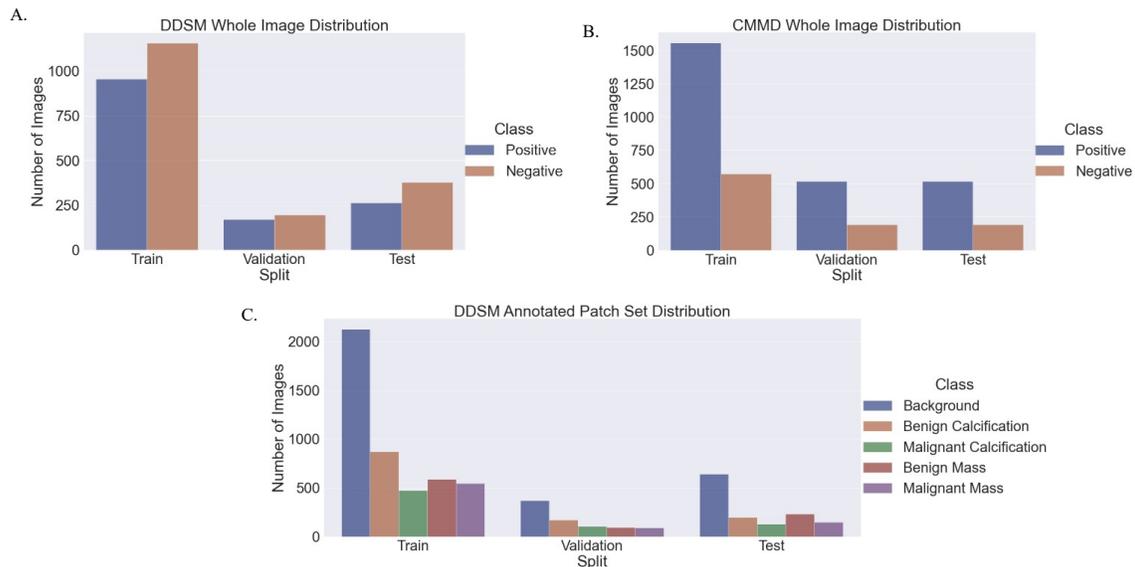

Supplemental Figure 1. Class distributions of A. DDSM whole mammograms, B. CMMD whole mammograms, and C. our annotated patch dataset drawn from DDSM.

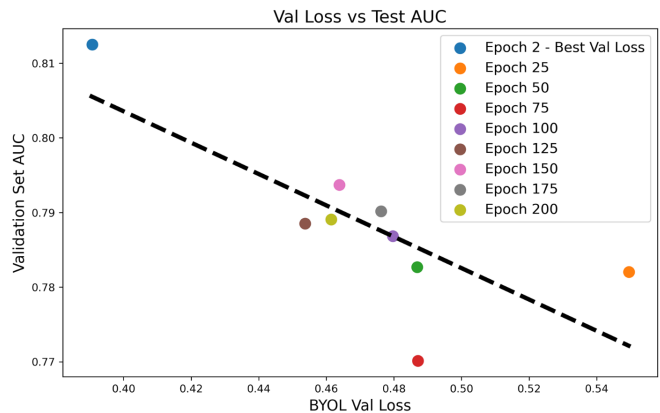

Supplemental Figure 2. Whole image maximal validation set AUC during linear evaluation vs validation set pretraining loss. The dashed line indicates a linear regression between validation loss and validation set AUC. There is a moderately strong correlation between the two (R = -0.79).

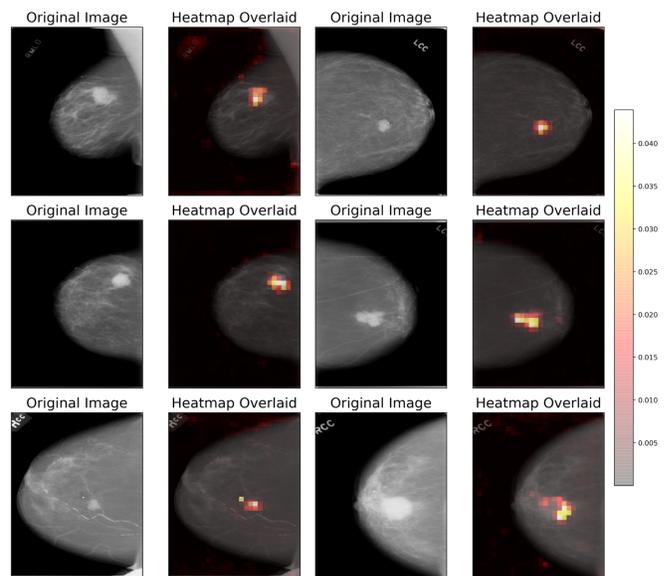

Supplemental Figure 3. Sample attention heatmaps from 6 mammograms in our training set. Attention was trained on top of a frozen encoder network pretrained with BYOL on the 96x96 tiled patch set. The colorbar corresponds to probabilities generated by the softmax layer in MIL attention – larger numbers indicate a larger weight assigned to the respective patch.

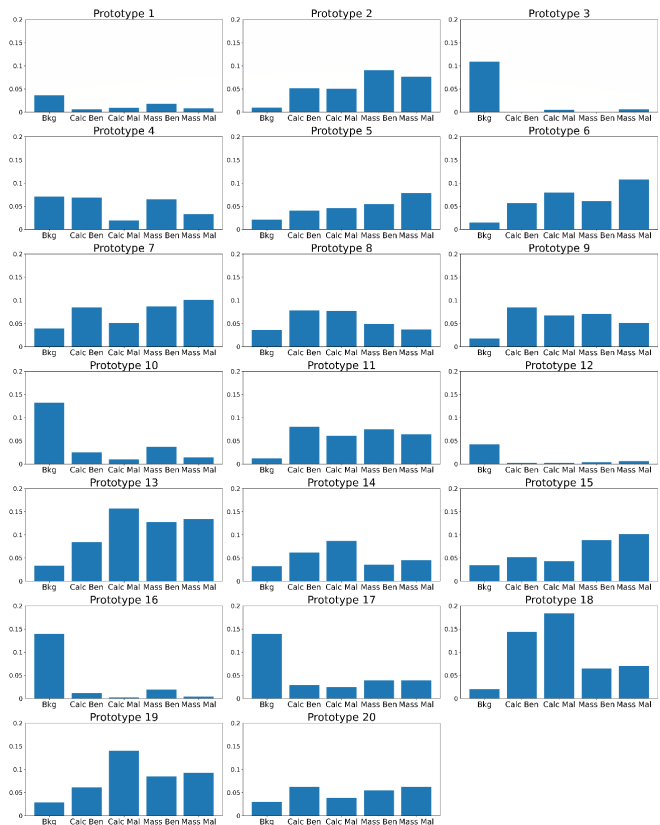

Supplemental Figure 4. Class distribution of the annotated patch set assigned with the SWaV model learned on the 96x96 tiled patches. Values are normalized to the number of images per class in the annotated patch set.

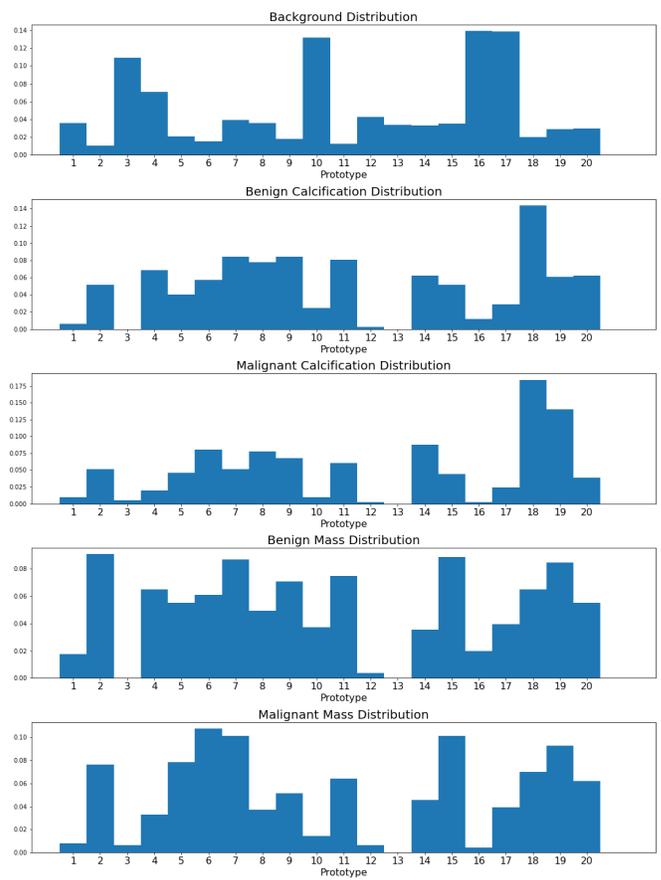

Supplemental Figure 5. Distribution of the 20 prototypes for each annotated class. Values are normalized to proportions.